\documentclass{mem}
\usepackage{natbib}\usepackage{txfonts}\usepackage{balance}
\usepackage{graphicx}
\usepackage[a4paper]{hyperref}
\idline{75}{282}
\begin{document}
\def\teff{$T\rm_{eff }$}
\def\kms{$\mathrm {km s}^{-1}$}

\title{
Fueling QSOs: The Relevance of Mergers}

\author{
N. \,Bennert\inst{1} 
\and G. \,Canalizo\inst{1}
\and 
B. \,Jungwiert\inst{2}
\and A. \,Stockton\inst{3}
\and F. Schweizer\inst{4}
\and C. Peng\inst{5}
\and M. Lacy\inst{6}
}

  \offprints{N. Bennert}

\institute{
Institute of Geophysics and 
Planetary Physics, UC Riverside, CA 92521, USA
\email{nicola.bennert@ucr.edu}
\and
Astronomical Institute, Academy of Sciences,
Bo{\v c}n\'\i\ II 1401, 141 31 Prague 4, Czech Republic
\and
Institute for Astronomy, University of Hawaii, 2680 Woodlawn Dr., 
Honolulu, HI 96822, USA
\and
Carnegie Observatories, 
813 Santa Barbara Street, Pasadena, CA 91101, USA
\and
NRC Herzberg Institute of Astrophysics, 5071 West Saanich
Road, Victoria, British Columbia, Canada V9E 2E7
\and
Spitzer Science Center, 
California Institute of Technology, Pasadena, CA 91125, USA
}

\authorrunning{Bennert et al.}
\titlerunning{Fueling AGNs: The Relevance of Mergers}

\abstract{
To study the relevance of mergers for
the fueling of QSOs, we are currently conducting an
HST imaging campaign of a sample of QSO host galaxies
classified as ellipticals in the literature. 
Here, we present results from a study of 
the first five QSO host galaxies imaged with HST/ACS.
For the majority of objects,
strong signs of interactions
such as tidal tails, shells, and other fine structure are revealed. 
We estimate the nature and
age of the merger by comparing the images with numerical simulations.
The merger ages range between a few hundred Myr up to a Gyr. These timescales
are comparable to starburst ages in the QSO hosts previously
inferred from Keck spectroscopy, but longer than theoretical
estimates of AGN duty cycles. A possible scenario emerging 
from our results is that most QSO host galaxies experienced mergers
with accompanying starbursts but that the activity is triggered with
a delay of several hundreds Myr after the merger.
To probe whether there is indeed a causal connection
between the merger and the QSO activity, we study a control
sample of inactive ellipticals.
Our preliminary results do not reveal comparable fine structure.
}
\maketitle{}

\section{Introduction}
The ubiquity of supermassive black holes (BHs) in the center of galaxies shows that more 
than the mere presence of a massive BH is needed to trigger the activity observed in Active 
Galactic Nuclei (AGNs).
Moreover, the steep evolution of activity with redshift indicates that the accretion
onto the BH must have been more common in the earlier universe and thus also the triggering
mechanism.
Mergers have been suggested to induce the sudden
inflow of gas to the center needed to feed the AGN
\citep[e.g.][]{too72,sto82,san88}.

A close connection between mergers and AGN activity
has been found for ultra-luminous infra-red galaxies (ULIRGs).
Observations of these galaxies are consistent
with an evolutionary scheme in which (at least some) ULIRGs
form through mergers of gas-rich galaxies and
represent the initial dust-enshrouded
stage in the evolution of optically selected QSOs
\citep[e.g.][]{san88,can01,vei06}. Finally, the QSO hosts
may end up as inactive elliptical galaxies.

However, a general connection between mergers and AGN activity is 
still being debated. While there is little doubt that mergers
are helpful to provide the gas and remove angular momentum, 
they are certainly not sufficient,
considering the numerous examples of inactive interacting galaxies.
Also, mergers may be necessary for QSOs only. For their low-luminosity
cousins, Seyfert galaxies, often residing in spiral host galaxies,
there is little direct evidence for unusually high rates of 
interaction \citep[e.g.][]{mal98}. Secular evolution
through processes such as bar instabilities may be the dominant
effect in the evolution of these galaxies \citep[e.g.][]{com06}.
But even for QSOs, the role of mergers for the activity
remains unclear. Recent high-resolution imaging studies
showed that many QSOs reside in elliptical hosts \citep[e.g.][]{dis95,bah97,flo04}.
From HST/WFPC2 images of
33 AGNs (radio-loud QSOs, radio-quiet QSOs, and radio galaxies)
at a redshift of $z \simeq 0.2$,
\citet{dun03} concluded that ``for nuclear luminosities $M_V$ $<$ -23.5, the hosts
of both radio-loud {\em and} radio-quiet AGN are virtually all massive
elliptical galaxies with basic properties that are indistinguishable
from those of quiescent, evolved, low-redshift ellipticals of comparable mass''.

To study the relevance of mergers for the
fueling of classical QSOs, we are currently conducting an
HST imaging campaign of a sample of QSO host galaxies
classified as ellipticals by \citet{dun03}. 
Here we present results from a pilot study of 
five QSO host galaxies ($z \simeq 0.2$) using
very deep  (5 orbits) HST/ACS images (F606W).

\section{Results and Discussion}
\begin{figure*}[t!]
\begin{center}
{\includegraphics[clip=true,scale=0.5]{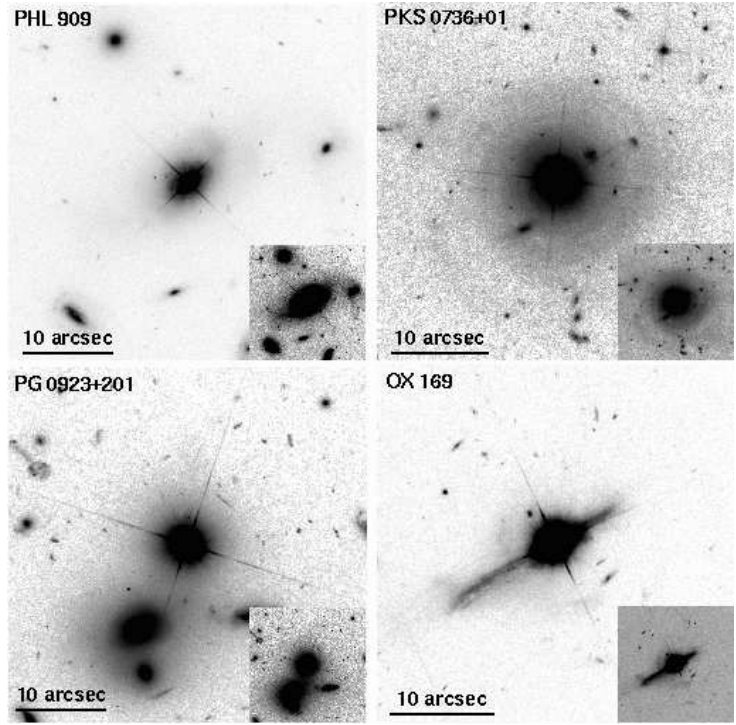}}
\caption{\footnotesize
Deep HST/ACS images of four of the five early-type QSO host galaxies \citep{ben08}.
North is up, east is to the left.
{\bf PHL\,909} ({\it top left}): A ring-like structure is seen
around the QSO nucleus. Diffuse outer material form another ring and tidal
tails to both sides of the host.
{\bf PKS\,0736+01} ({\it top right}): A large ($r$ $\sim$ 50\,kpc) but faint spiral-like
structure surrounds the QSO. The irregular structure and
changes in pitch angle may indicate spatial wrapping
of material from a (minor) merger event rather than a spiral disk seen face on.
{\bf PG\,0923+201} ({\it bottom left}): No fine structure can be seen in the host galaxy,
but it lies in an environment with several interacting companions. 
{\bf OX\,169} ({\it bottom right}):
The extended linear structure is likely a tidal tail seen nearly edge on. Note the extended shell-like
features east of the nucleus.
}\label{fourqsos}
\end{center}
\end{figure*}

\begin{figure*}[t!]
\begin{center}
{\includegraphics[clip=true,scale=0.5]{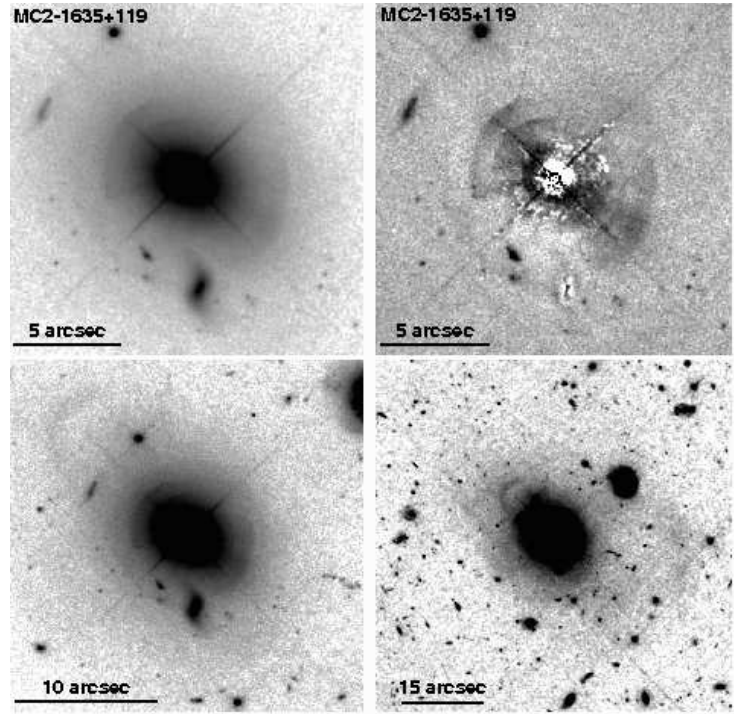}}
\caption{\footnotesize
The same as in Fig.~\ref{fourqsos} for MC2\,1635+119 \citep{can07}.
Spectacular interleaved shells occur at $r$ $\sim$ 5-12\,kpc ({\it top left}).
The shell structure is seen more prominently
when subtracting a PSF+host galaxy model as fitted by GALFIT 
(\citealt{pen02}; {\it top right}).
An arc-like feature extends out to $\sim$ 32\,kpc ({\it bottom left}).
An even larger, faint and diffuse structure 
can be seen $\sim$65 kpc west of the center ({\it bottom right}).
}\label{mc}
\end{center}
\end{figure*}

For four of the five QSOs,
our images reveal dramatic signs of interactions
such as tidal tails, shells, and other fine structure in the
hosts (Figs.~\ref{fourqsos},\ref{mc}),
suggesting that a large fraction of QSO host
galaxies may have experienced a relatively recent merger event \citep{ben08}.
 
One spectacular example of regular inner shell structure is MC2 1635+119
\citep{can07}. 
In $N$-body simulations, the observed shells can be produced in a 
minor merger event. Assuming
that the outermost shell is at 12.5 kpc (Fig.~\ref{mc}),
we estimate a merger age of $\sim$30-400 Myr,
depending on the type of profile of the giant elliptical,
its effective radius, and the amount of dark matter.
Taking into account a larger tidal feature at 65 kpc
(Fig.~\ref{mc}) that may
be an older shell formed during the same encounter,
the merger age can be up to 1.7 Gyr \citep{can07}.
However, while the inner shell structure can be produced
by a radial minor merger,
we cannot exclude other scenarios such as a major merger. 
Indeed, the total light contribution from the shells ($\sim$6\%) and extended
structures are more indicative of a major merger. In this
case, the inner shell structure might have been formed by material ``raining''
back into the central regions of the merger remnant \citep{can07}.

For all objects, deep Keck spectroscopy revealed major starburst episodes 
($\sim$1-2 Gyr; \citealt{can06,can07}).
These timescales are comparable to the merger ages,
but significantly longer than theoretical estimates for QSO duty cycles \citep[e.g.][]{yu02}.
Our results indicate that while most QSO host galaxies experienced mergers
with accompanying starburst, there is a time delay
of several hundred Myr between the tidal interaction and the actual
fueling of the central BH.
This is in agreement with recent
hydrodynamic simulations \citep[e.g.][]{spr05,hop07}.

\section{Outlook}
The question remains whether the QSO host galaxies are truly
distinct from inactive ellipticals or whether we can find 
similar fine structure hinting a recent merger event.
To address this question, we selected a control sample of elliptical galaxies from the HST/ACS archive.
So far, none of the ellipticals has shown the spectacular fine structure found in the 
QSO hosts, although some have apparent companions.

Also, we are currently studying 14 more QSO host galaxies imaged with HST/WFPC2.
Preliminary results show evidence for mergers in at least some of
these QSO hosts as well.

\begin{acknowledgements}
This work was supported in part under proposals GO-10421 and AR-10941 by 
NASA through a grant from the Space Telescope Science Institute, which is 
operated by the Association of Universities for Research in Astronomy, Inc., 
under contract NAS5-26555. Additional support was provided by the National 
Science Foundation, under grant number AST 0507450. NB is grateful for support
from the American Astronomical Society and the 
National Science Foundation in the form of an International Travel Grant,
enabling her to attend the conference. NB acknowledges support by
'RadioNet', 
funded under the European Commission's Sixth Framework Programme.
BJ acknowledges support by the Grant No. LC06014 of the Czech Ministry
of Education and by the Research Plan No. AV0Z10030501 of the Academy
of Sciences of the Czech Republic.
\end{acknowledgements}

\bibliographystyle{aa}

\begin{thebibliography}{}
\bibitem[Bahcall et al.(1997)]{bah97} Bahcall, J. N. et al. 1997, ApJ, 479, 642
\bibitem[Bennert et al.(2008)]{ben08} Bennert, N. et al. 2008, \apj, 677, 846
\bibitem[Canalizo \& Stockton(2001)]{can01} Canalizo, G. \& Stockton, A. 2001, \apj, 555, 719
\bibitem[Canalizo et al.(2006)]{can06} Canalizo, G. et al. 2006,
New Astron.Rev., 50, 650
\bibitem[Canalizo et al.(2007)]{can07} Canalizo, G. et al. 2007, \apj, 669, 801
\bibitem[Combes(2006)]{com06} Combes, F. 2006, RMxAC, 26, 131
\bibitem[Disney et al.(1995)]{dis95} Disney, M. J. et al. 1995, Nature, 376, 150
\bibitem[Dunlop et al.(2003)]{dun03} 	
Dunlop, J. S. et al. 2003, \mnras, 340, 1095 
\bibitem[Floyd et al.(2004)]{flo04} Floyd, D. J. E. et al.  2004, MNRAS,
355, 196
\bibitem[Hopkins et al.(2007)]{hop07} Hopkins, P. F. et al. 2007, ApJ, 662, 110
\bibitem[Malkan, Gorjian, \& Raymond(1998)]{mal98} Malkan, M. A., Gorjian, V., \& Raymond, T. 1998, ApJS, 117, 25
\bibitem[Peng et al.(2002)]{pen02} 
Peng, C. Y. et al. 2002, AJ, 124, 266
\bibitem[Sanders \& Mirabel(1996)]{san96} Sanders, D. B., \& Mirabel, I. F. 1996,
ARA\&A, 34, 749
\bibitem[Sanders et al.(1988)]{san88} Sanders, D. B. et al. 1988, ApJ, 325, 74
\bibitem[Springel, Di Matteo \& Hernquist(2005)]{spr05} Springel, V., Di Matteo, T.,
\& Hernquist, L. 2005, MNRAS, 361, 776
\bibitem[Stockton(1982)]{sto82} Stockton, A. 1982, ApJ, 257, 33
\bibitem[Toomre \& Toomre(1972)]{too72} Toomre, A., \& Toomre, J. 1972, ApJ, 178, 623
\bibitem[Veilleux et al.(2006)]{vei06} Veilleux, S. et al. 2006, ApJ, 643, 707
\bibitem[Yu \& Tremaine(2002)]{yu02} Yu, Q., \& Tremaine, S. 2002, MNRAS, 335, 965
\end{thebibliography}

\end{document}